# Temporal nonlinear dynamics of plasmon-solitons, a Duffing oscillator-based approach

**Morteza A. Sharif,**[a,*]

[a]Urmia University of Tehnology, Faculty of Electrical engineering, Optics and Laser engineering group, Band road, Urmia, Iran, 5716617165

**Abstract**. This paper deals with the temporal nonlinear dynamics of plasmon-solitons in a plasmonic waveguide. Duffing equation is recognized as the temporal part of the nonlinear amplitude equation governing the plasmonic waveguide. It is shown that Duffing oscillator waveforms stand for the temporal nonlinear dynamics of plasmon-soliton waves. The exchange of Lorentz-type bright and dark solitons' energies gives rise to a Fano resonance. It is thus shown that the interaction of solitons and the formation of plasmon-solitons is inherently nonlinear. It is accordingly indicated that the nonlinear modulation of the plasmon-solitons is achievable via tuning the nonlinearity of the plasmonic waveguide.

*****Morteza A. Sharif**, E-mail: m.abdolahisharif@ee.uut.ac.ir

## 1 Introduction

Duffing equation is a mathematical tool to model a nonlinear dynamical system with damped and driven oscillations. Although Duffing equation has been generally used for the mechanical systems, a variety of studies has addressed it in the nonlinear optical systems. K. Senthilnathan, et al developed an inharmonic oscillator equation for bright solitons in photorefractive media. Then, they showed that the equation can be reduced to undamped/unforced Duffing equation [1]. E Babourina-Brooks, et al. determined the noise response of a quantum nanomechanical resonator using a Duffing oscillator-based model due to the dynamical equivalency [2]. Heung-Ryoul Noh indicated that the Duffing oscillator can be realized in an intensity-modulated magneto-optical trap [3]. Considering the Kerr nonlinearity, Wiktor Walasik, et al. modeled the field profiles of plasmon-soliton waves in planar slot waveguide using the Jacobi elliptic special functions [4]. Jacobi elliptic functions are indeed the stationary solutions of the Duffing equation [5]. M. Scalora, et al. employed a Duffing oscillator-based model to investigate the third harmonic generation in metal nanostructures [6]. We have recently proposed Duffing oscillator model for delineation of dynamical states and chaotic maps in a surface plasmon laser [7].

Surface Plasmons (SPs) are intrinsically electrons that oscillate at an interface (between two materials) or a surface (like a quantum dot surface or a 2D material). Their physical properties are closer to the bosons implying that SPs can intensely be localized beyond the diffraction limit of the light [8]. This is essential for the next generation nanophotonic devices[9-20]. One important drawback is the propagation length which is confined to several micrometers resulting from the damping nature of SPs. Plasmonic waveguides have been thus investigated and developed for the two purposes of increasing the propagation length and keeping the strong localization [19-23]. The literature is almost full of the studies investigating the linear dynamics to derive the plasmon modes. However, some recent studies have shown that the latter is insufficient due to the large nonlinearity provided by the plasmonic waveguides' nanostructure. Sergey Mikhailov showed that the influence of nonlinearity on plasmon resonances at special frequencies are the broadening and reducing the propagation length [24]. On the side, Wiktor Walasik, et al. considered the Kerr nonlinearity for the waveguide to obtain the stationary solutions [4]. As well, we found that the unstable solutions and breather modes can be resulted for a plasmonic waveguide in consequence of the large nonlinear response [25].

In this article, Duffing equation is re-considered to derive temporal stationary solutions in a plasmonic waveguide. It is shown that the nonlinearity is responsible for a Self-Amplitude Modulation (SAM) effect. It is also indicated that Lorentz-type soliton as well as Fano resonance can be obtained as the solitary solution if the absorption and nonlinear coefficients with complex values are proportionally tuned. The ratio of nonlinear to absorption coefficient is recognized as a determining factor for modulating the temporal dynamics of the plasmon-soliton waves.

## 2 Theory

The nonlinear equation to describe the spatiotemporal SPs' amplitude $\Psi$ in a plasmonic waveguide is given by Eq.(1)[37].

$$i\left(\frac{\partial \Psi}{\partial z} + D^{-1}\frac{\partial \Psi}{\partial t}\right) - \frac{k'}{2}\frac{\partial^2 \Psi}{\partial t^2} + \gamma|\Psi|^2 A + i\eta\Psi = \delta\cos(\omega_d t), \tag{1}$$

where $\gamma$ is the nonlinear coefficient; $\eta$ is the linear absorption coefficient which characterizes the smoothness of oscillation; $\delta$ and $\omega_d$ are respectively, the amplitude and angular frequency of the driving force. $D^{-1} = \left(\frac{dk}{d\omega}\right)_{\omega=\omega_0} = (v_g)^{-1}_{\omega=\omega_0}$; $k$ is the wavenumber; $\omega$ is the angular frequency; $v_g = \frac{d\omega}{dk}$ is the group velocity; $k'' = \left(\frac{d^2k}{d\omega^2}\right)_{\omega=\omega_0}$. Assuming $\delta=0$, Eq.(1) can be separable and the spatiotemporal amplitude can be written as $\Psi(z,t) = Z(z)A(t)$. In order to obtain $A(t)$, one should first assume the temporal part of Eq.(1) as in the form of Eq.(2).

$$\frac{k''}{2}\frac{d^2 A}{dt^2} + \gamma|A|^2 A + i\eta A = 0, \tag{2}$$

which can be further reduced to the form given in Eq.(3).

$$\frac{d^2 A}{dt^2} + \gamma'|A|^2 A + \eta' A = 0, \tag{3}$$

where it has been assumed that $\gamma' = \frac{2\gamma}{k''}$ and $\eta' = \frac{-2i\eta}{k''}$. Considering $A$ as an observable wave function, time-energy uncertainty principal will be written as Eq.(4). From the viewpoint of quantum theory, representation of $A$ vs. the time also implicates the system energy eigenvalues explainable by Eq.(4) [38].

$$\sigma_E \frac{\sigma_A}{\frac{d\langle A\rangle}{dt}} \geq \frac{\hbar}{2}, \tag{4}$$

where $\sigma_E$ and $\sigma_A$ are respectively, the energy and wave function eigenvalues; $\langle A \rangle$ is the expectation value and $\hbar$ is the reduced Planck's constant. Analytical solution of Eq.(3) can be given by Eq.(5) [5].

$$A(t) = c_1 \, \text{cn}\left[ \left(\eta' + c_1^2 \gamma'\right)^{\frac{1}{2}} t + c_2, \left( \frac{c_1^2 \gamma'}{2(\eta' + c_1^2 \gamma')} \right)^{\frac{1}{2}} \right], \qquad (5)$$

where cn, sn and dn are the Jacobi elliptic functions and $c_1$ and $c_2$ are to be obtained from the initial conditions given in Eq.(6) [5].

$$\begin{cases} c_1 \, \text{cn}\left( c_2, \sqrt{\dfrac{c_1^2 \gamma'}{2(\eta' + c_1^2 \gamma')}} \right) = A_0 \\ -\sqrt{\eta' + c_1^2 \gamma'} \, \text{sn}\left( c_2, \sqrt{\dfrac{c_1^2 \gamma'}{2(\eta' + c_1^2 \gamma')}} \right) \text{dn}\left( c_2, \sqrt{\dfrac{c_1^2 \gamma'}{2(\eta' + c_1^2 \gamma')}} \right) = A_{z_0} \end{cases}, \qquad (6)$$

where $A_0$ and $A_{z_0}$ are the initial and boundary values of the wave amplitude $A$ at $z = 0$ and $z = z_0$. Answers given in Eq.(5) are stationary solutions prior to unstable states. Transition to instability is beyond this study and has been investigated in our previous study [7]. If $\gamma' = \dfrac{-2\eta'}{A_0^2}$, solution of Eq.(5) will be reduced to Eq.(7)[5].

$$A(t) = A_0 \, \text{sech}\left( \sqrt{\eta'} \, t \right). \qquad (7)$$

Solitary solution given in Eq.(7) stands for negative values of $\gamma'$ which is thus fulfilled for the real values of $\sqrt{\eta}$. If otherwise, $\sqrt{\eta}$ is generally a complex parameter, the solution of Eq.(7) will be accordingly modified.

## 3    Simulation results and discussion

*Ansatz I. Real positive absorption coefficient*

Fig.1 shows the solitary solution given in Eq.(7) for different values of the nonlinear coefficient $\gamma'$. As the latter increases, the soliton width decreases.

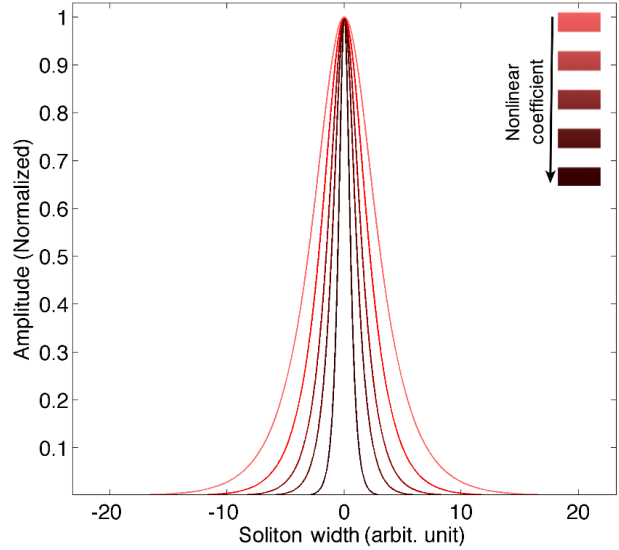

Fig 1. soliton width for different values of nonlinear coefficient

On the other hand, general solutions of Eq.(6) are very sensitive to the ratio of the nonlinear coefficient to absorption coefficient i.e. $\frac{\gamma'}{\eta'}$. Oscillatory behavior in Fig,2 unveils a SAM. This modulation feature is to be broadened for the lower nonlinear coefficients as the nonlinear coefficient increases (Fig.2(a)) and vice versa (Fig.2(b)). The broadening also means increasing the number of pulses per each wave envelop. This result may also reveal the broadening of the plamonic waves for special considered frequencies as indicated by Sergey Mikhailov et al [24].

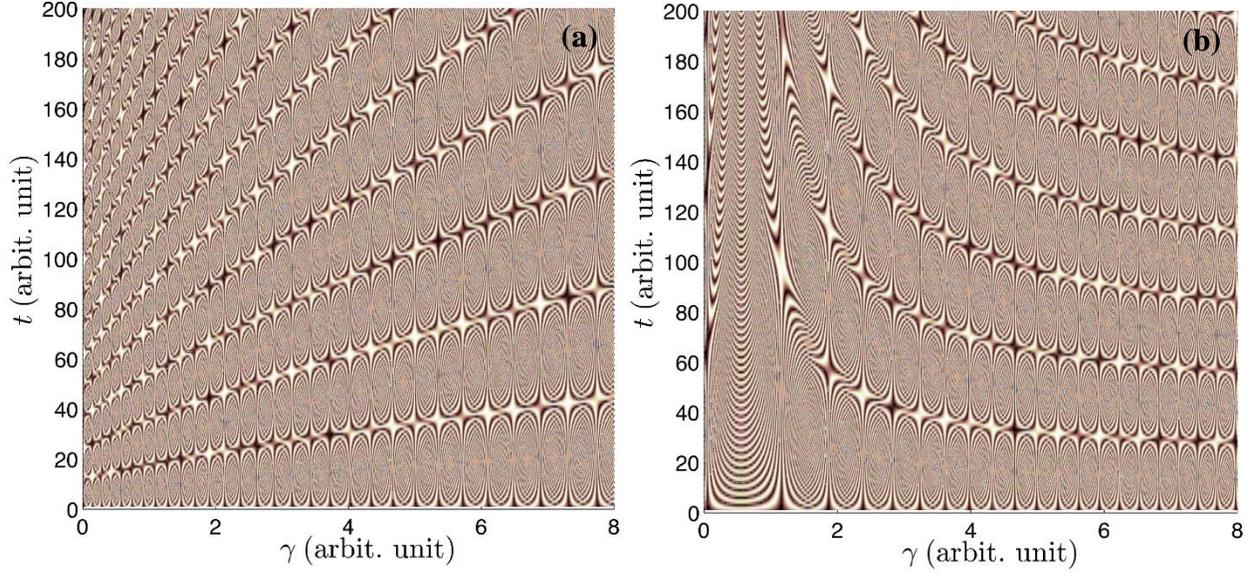

Fig 2. Oscillatory Duffing waveforms for different values of nonlinear coefficient $\gamma$ when (a) $\gamma/\eta \sim 10^{-8}$; (b) $\gamma/\eta \sim 10^{-5}$.

Time evolution of the Duffing oscillator is depicted in Fig. 3 for different values of $\frac{\gamma'}{\eta'}$. The amplitude evolves from the single periodic state (Fig. 3(a)) to quasi-period state (Fig.3(b), (c) & (d)). The phase and group velocities are interpreted to be identical in Fig.3(a). Then, a difference in phase and group velocities leads to SAM as shown in Fig.3(b), (c) & (d). Modulation depth raises up to 0.89 for Fig.3(c) and 0.97 for Fig.3(d) while the modulation rate significantly increases.

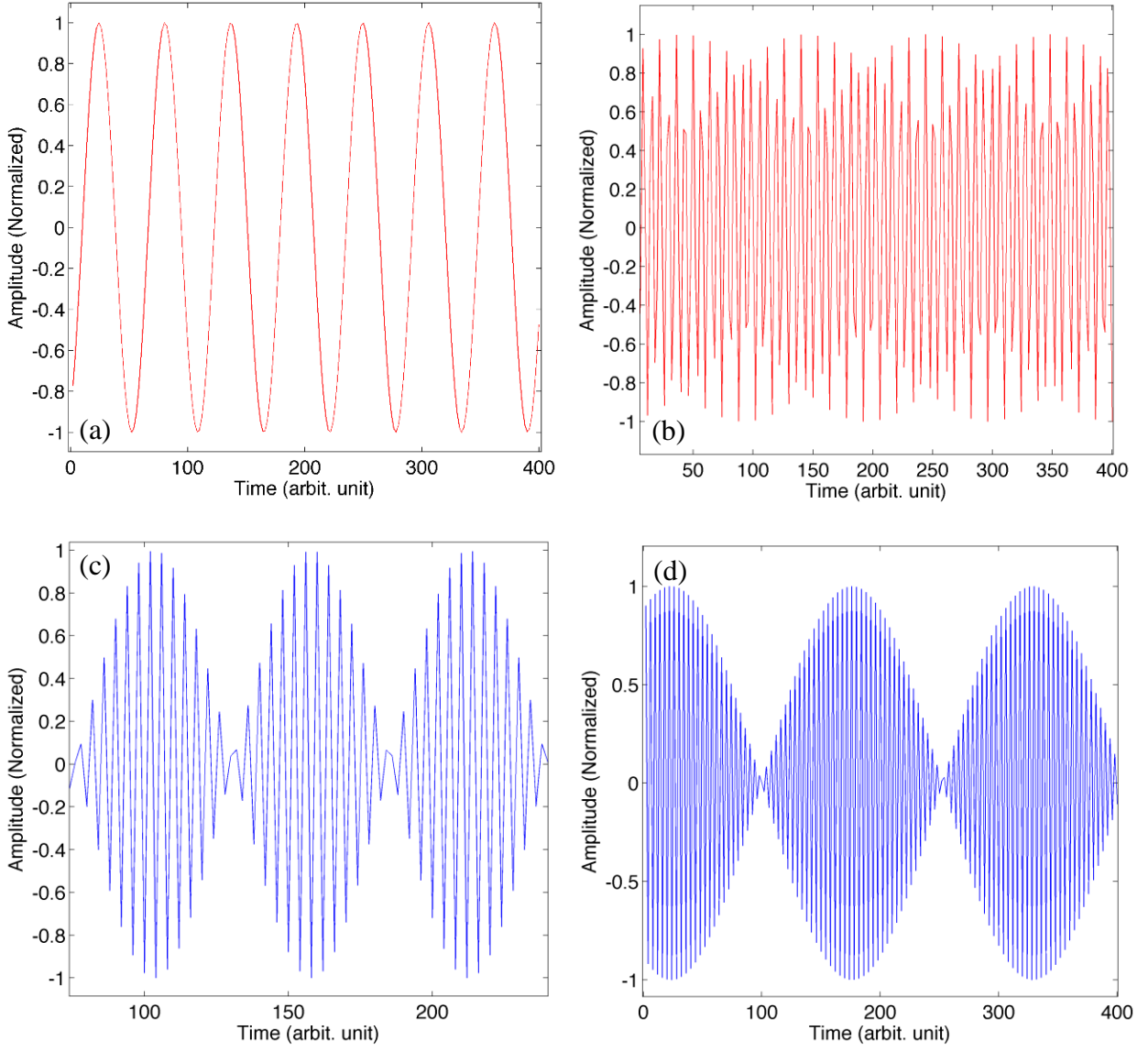

Fig 3. Time evolution of Duffing waveforms for (a) $\gamma'_1/\eta'_1 = 1.675 \times 10^{-6}$; (b) $\gamma'_2/\eta'_2 = 3.75 \times 10^{-5}$, $\gamma'_2 = 0.22\, \gamma'_1$, $\eta'_2 = 1/100\, \eta'_1$; (c) $\gamma'_3/\eta'_3 = 7.5 \times 10^{-6}$, $\gamma'_3 = 1/5\gamma'_2$, $\eta'_3 = \eta'_2$ and (d) $\gamma'_4/\eta'_4 = 7.5 \times 10^{-8}$, $\gamma'_4 = 1/5\gamma'_3$, $\eta'_4 = \eta'_3$.

*Ansatz II. Complex absorption coefficient, plasmon-soliton waves*

If $\gamma' = \dfrac{-2\eta'}{A_0^2}$, a solitary shape can be obtained as depicted in Fig.4 for different values of the nonlinear coefficient $\gamma$ and initial amplitude $A_0$. Although the results imply the formation of a flat-top solitary shape (in comparison with Fig.1), larger nonlinear coefficients as well as the lower initial amplitudes (Fig.4(a) compared to Fig.(b)) lead to decreasing the soliton width.

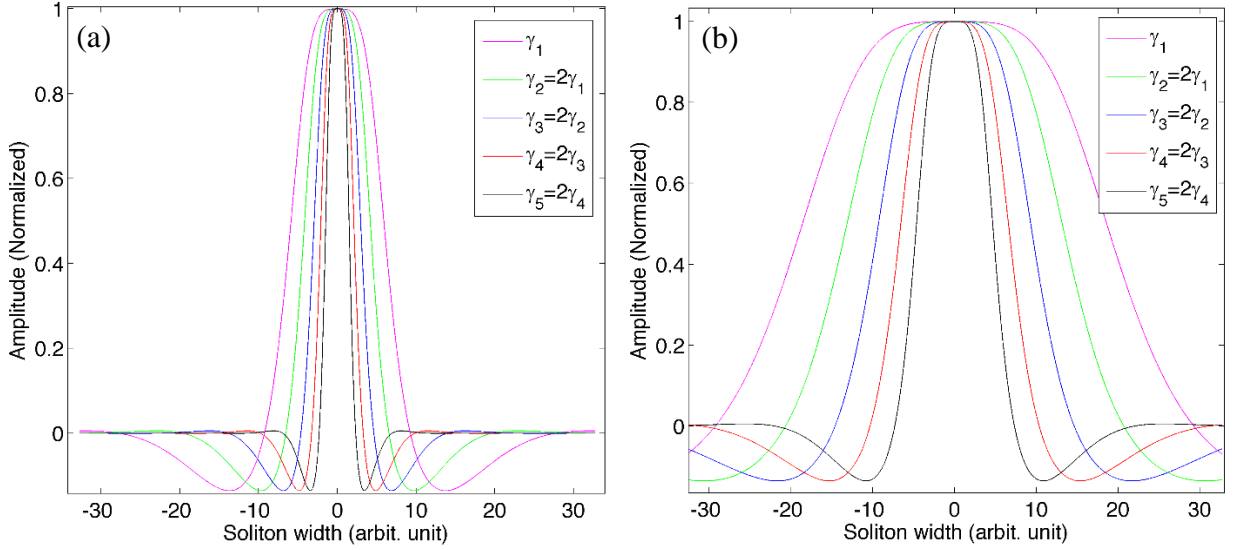

Fig 4. (a) Soliton width for different values of nonlinear coefficient, (b) the same for 10 times higher initial amplitude $A_0$.

In addition, temporal dynamics obtained for some assumed values of $\frac{\gamma'}{\eta'}$ are shown in Fig.5 in accordance with the modified form of the general solutions in Eq. (6) provided for the complex values considered for the absorption coefficient. The results include the Lorentz-type solitons [40] (Fig.5(b) & Fig.5(d)), Fano resonance [41-43] (Fig.5(c)) and a combination of both (Fig.5(a) & Fig.5(d)).

Spatiotemporal solutions of Eq.(1) can be obtained in accordance with the wave function $\Psi = A(t) Z(z)$ if one considers that the spatial part to be conventionally given by Eq.(8).

$$Z(z) \propto \begin{cases} e^{-X_\perp} & X_\perp > 0 \\ e^{X_\perp} & X_\perp < 0 \end{cases}, \qquad (8)$$

where $X_\perp$ is any coordinate perpendicular to the propagation direction[5]. Some spatiotemporal plasmon-soliton modes are shown in Fig.6 for different values assumed for $\frac{\gamma'}{\eta'}$.

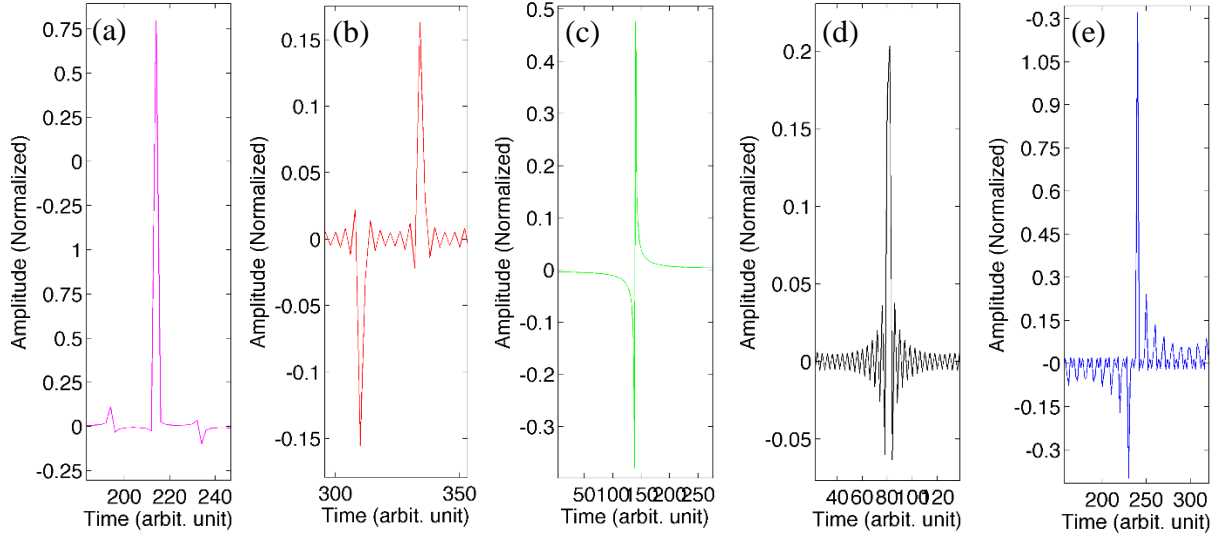

Fig 5. Time evolution of Duffing waveforms for (a) $\gamma'_1/\eta'_1=0.6\times10^{-4}$; (b) $\gamma'_2/\eta'_2=2.5\times10^{-7}$, $\gamma'_2=0.01\,\gamma'_1$, $\eta'_2=10\,\eta'_1$; (c) $\gamma'_3/\eta'_3=5\times10^{-6}$, $\gamma'_3=2\gamma'_2$, $\eta'_3=0.1\,\eta'_2$, (d) $\gamma'_4/\eta'_4=1\times10^{-6}$, $\gamma'_4=2\gamma'_3$, $\eta'_4=\eta'_3$ and (e) $\gamma'_5/\eta'_5=8\times10^{-5}$, $\gamma'_5=8\gamma'_4$, $\eta'_5=0.1\,\eta'_4$.

*Discussion*

In real systems (like the plasmonic waveguides), the values of the nonlinear and absorption coefficients read the relations $\gamma \propto \chi^{(3)}$ and $\eta \propto \chi^{(1)}$ where $\chi^{(3)}$ and $\chi^{(1)}$ are the third order and linear susceptibility of the system structure respectively [44]. If $\chi^{(1)}$ is reduced to unity, one can write $\chi^{(3)} \simeq \left(4\pi\varepsilon_0\right)^6 \hbar^8 / m^4 e^{10}$ [44]. Even so, one may notice that the ratio of $\dfrac{\gamma'}{\eta'}$ can be varied depending on the characteristics of the waveguide such as the wavelength, number of 2D layers, Fermi energy, driving voltage/light intensity, etc. A variety of studies has shown the Fano resonance and Lorentz-type solitons as prevalent dynamical regime of the plasmonic waveguide in the time domain [41-43]. In corroboration, the results obtained for *Ansatz II* can be truly assigned to the temporal nonlinear dynamics of the plasmon-solitons. Energy exchange between the dark and bright modes causes a Fano resonance. The results suggests that the analytical solutions of Duffing equation (Eq.(3)) can describe the stationary states in a plasmonic waveguide. One other important issue is the feasibility for the nonlinear modulation of the plasmon-soliton waves by tuning the value $\frac{\gamma'}{\eta'}$.

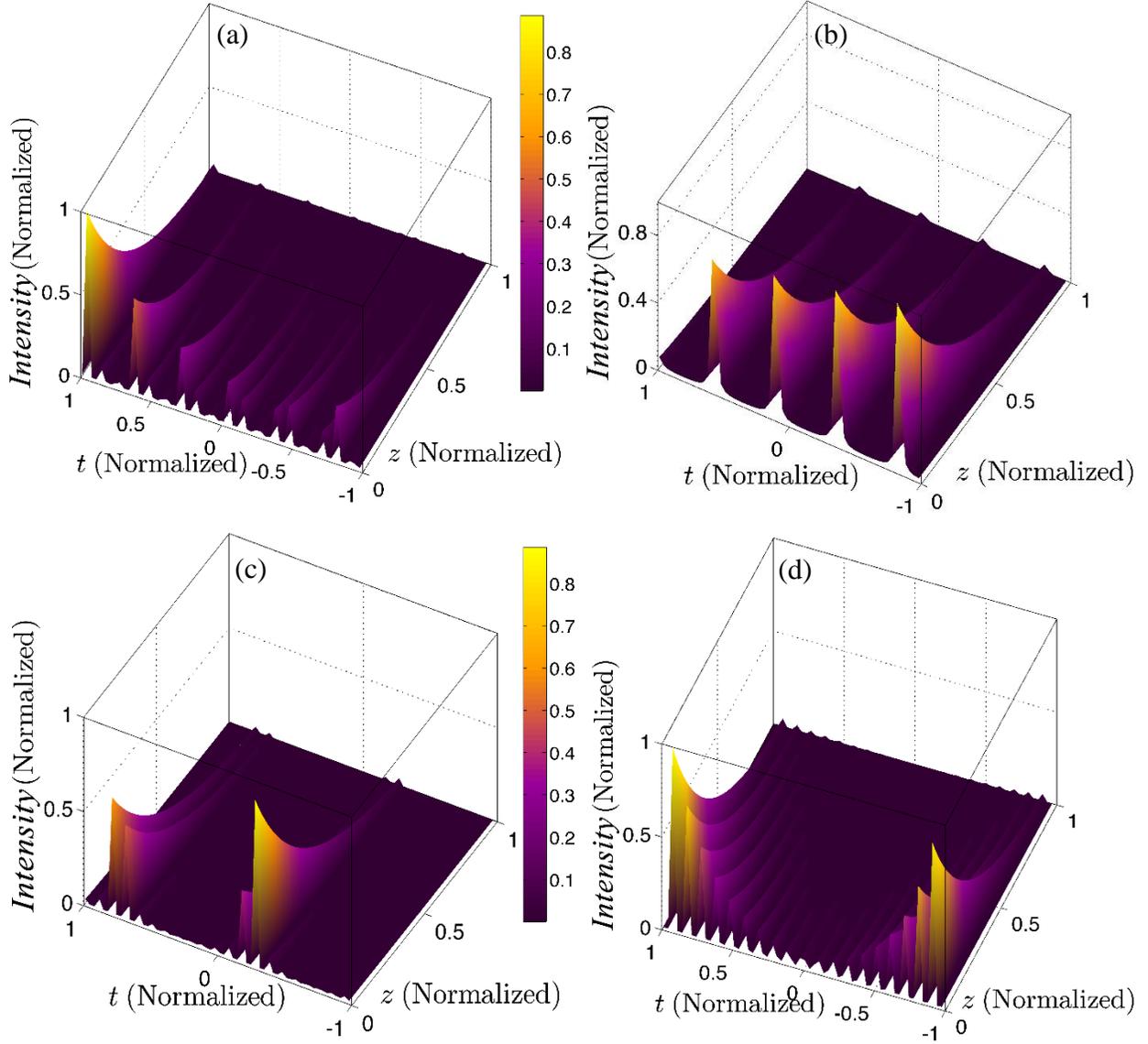

Fig 6. Spatiotemporal evolution of plasmon-soliton modes for (a) $\gamma'_1/\eta'_1=0.8\times10^{-4}$; (b) $\gamma'_2/\eta'_2=0.9\times10^{-3}$, $\gamma'_2=11.25\ \gamma'_1$, $\eta'_2=\eta'_1$; (c) $\gamma'_3/\eta'_3=1.4\times10^{-4}$, $\gamma'_3=11\ \gamma'_2$, $\eta'_3=7.6\times10^3\ \eta'_2$; (d) $\gamma'_4/\eta'_4=8\times10^{-7}$, $\gamma'_4=8\times10^{2}\gamma'_3$. $\eta'_4=1.3\times10^5\ \eta'_3$.

## 4    Conclusion

Solitary solutions of the temporal nonlinear dynamics in a plasmonic waveguide has been derived using the analytical solution of the Duffing equation as the temporal part of the spatiotemporal nonlinear amplitude equation. It has been proposed that the Duffing oscillator waveforms obtained for the complex values of the absorption and nonlinear coefficients stand for the temporal nonlinear dynamics of plasmon-soliton modes. It has been shown that the ratio of the nonlinear to absorption coefficient is an effective factor for determining the plasmon-soliton modes. Accordingly, tuning the absorption and nonlinearity is suggested for the nonlinear modulation of plasmon-soliton waves. The results are thus useful for effective design process of the plasmonic waveguides.

*References*

[1]  Senthilnathan, K., and K. Porsezian. "Bright and dark spatial solitons in coupled photorefractive media." *journal of modern optics* 51.3 (2004): 415-421..

[2]  Babourina-Brooks, E., A. Doherty, and G. J. Milburn. "Quantum noise in a nanomechanical duffing resonator." *New Journal of Physics* 10.10 (2008): 105020.

[3]  Moon, Geol, Wonho Jhe, and Heung-Ryoul Noh. "Duffing Oscillation in an Intensity-modulated Magneto-optical Trap: An Analytical Study for the (1+ 3) Atomic Energy Structure." *Journal of the Korean Physical Society* 58.5 (2011): 1105-1109.

[4]  Walasik, Wiktor, and Gilles Renversez. "Plasmon-soliton waves in planar slot waveguides. I. Modeling." *Physical Review A* 93.1 (2016): 013825.

[5]  Alvaro H. Salas, Jairo E. Castillo H.. Exact solution to Duffing equation and the pendulum equation. *Applied Mathematical Sciences*, Hikari, 2014, 8 (176), pp. 8781-8789. ⟨10.12988/ams.2014.44243⟩. ⟨hal-01356787⟩.

[6]  Scalora, M., et al. "Nonlinear Duffing oscillator model for third harmonic generation." *JOSA B* 32.10 (2015): 2129-2138.

[7]   Sharif, M.A., Ashabi, K. A Quasi-classical Model for Delineation of Dynamical States and Chaotic Maps in a Spaser. *Plasmonics* (2020). https://doi.org/10.1007/s11468-020-01269-6

[8]   Barnes, William L., Alain Dereux, and Thomas W. Ebbesen. "Surface plasmon subwavelength optics." *nature* 424.6950 (2003): 824-830.

[9]   Joly, Alan G., et al. "Surface Plasmon-Based Pulse Splitter and Polarization Multiplexer." *The journal of physical chemistry letters* 9.21 (2018): 6164-6168.

[10] Singh, Surya Prakash, Nilesh Kumar Tiwari, and M. Jaleel Akhtar. "Spoof Surface Plasmon-Based Coplanar Waveguide Sensor for Dielectric Sensing Applications." *IEEE Sensors Journal* 20.1 (2019): 193-201.

[11] Selvendran, S., et al. "A novel surface plasmon based photonic crystal fiber sensor." *Optical and Quantum Electronics* 52 (2020): 290.

[12] Zhang, Yusheng, and Zhanghua Han. "Experimental demonstration of spoof surface plasmon based THz antennas for huge electric field enhancement." *Plasmonics* 13.2 (2018): 531-535.

[13] Xu, Litu, et al. "Design of surface plasmon nanolaser based on MoS2." *Applied Sciences* 8.11 (2018): 2110.

[14] Li, Dongfang, and Domenico Pacifici. "Strong amplitude and phase modulation of optical spatial coherence with surface plasmon polaritons." *Science advances* 3.10 (2017): e1700133.

[15] Lee, Dong Hun, and Myung-Hyun Lee. "Efficient Experimental Design of a Long-Range Gapped Surface Plasmon Polariton Waveguide for Plasmonic Modulation Applications." *IEEE Photonics Journal* 11.1 (2019): 1-10.

[16] Lee, Ho Wai, et al. "Nanoscale plasmonic field-effect modulator." U.S. Patent No. 9,494,715. 15 Nov. 2016.

[17] Amin, Rubab, et al. "Heterogeneously Integrated ITO Plasmonic Mach-Zehnder Interferometric Modulator on SOI." *arXiv preprint arXiv:2007.15457* (2020).

[18] Stockman, Mark I., and David J. Bergman. "Surface plasmon amplification by stimulated emission of radiation (spaser)." U.S. Patent No. 7,569,188. 4 Aug. 2009.

[19] Klein, Matthew, et al. "2D semiconductor nonlinear plasmonic modulators." *Nature communications* 10.1 (2019): 1-7.

[20] Yuan, Dongdong, et al. "A multi-wavelength SPASER based on plasmonic tetramer cavity." *Journal of Optics* 21.11 (2019): 115001.

[21] Berini, Pierre. "Highlighting recent progress in long-range surface plasmon polaritons: guest editorial." *Advances in Optics and Photonics* 11.2 (2019): ED19-ED23.

[22] Jing, Jian-Ying, et al. "Long-range surface plasmon resonance and its sensing applications: A review." *Optics and Lasers in Engineering* 112 (2019): 103-118.

[23] Abdulhalim, Ibrahim. "Coupling configurations between extended surface electromagnetic waves and localized surface plasmons for ultrahigh field enhancement." *Nanophotonics* 7.12 (2018): 1891-1916.

[24] Mikhailov, Sergey A. "Influence of optical nonlinearities on plasma waves in graphene." *ACS Photonics* 4.12 (2017): 3018-3022.

[25] Sharif, Morteza A. "Spatio-temporal modulation instability of surface plasmon polaritons in graphene-dielectric heterostructure." *Physica E: Low-dimensional Systems and Nanostructures* 105 (2019): 174-181.

[26] Doi, Masaharu, et al. "Advanced LiNbO/sub 3/optical modulators for broadband optical communications." *IEEE Journal of selected topics in quantum electronics* 12.4 (2006): 745-750.

[27] Bulow, Jeffrey A. "Interferometric modulator for optical signal processing." U.S. Patent No. 5,315,370. 24 May 1994.

[28] Kuwabara, Nobuo, et al. "Development and analysis of electric field sensor using LiNbO/sub 3/optical modulator." *IEEE transactions on electromagnetic compatibility* 34.4 (1992): 391-396.

[29] Zhang, Yuxia, et al. "Broadband atomic-layer MoS 2 optical modulators for ultrafast pulse generations in the visible range." *Optics Letters* 42.3 (2017): 547-550.

[30] Merolla, Jean-Marc, et al. "Phase-modulation transmission system for quantum cryptography." *Optics letters* 24.2 (1999): 104-106.

[31] Urino, Yutaka, et al. "Demonstration of 12.5-Gbps optical interconnects integrated with lasers, optical splitters, optical modulators and photodetectors on a single silicon substrate." *Optics express* 20.26 (2012): B256-B263.

[32] Pham, Anh Tuan. "All-optical modulation and switching using a nonlinear-optical directional coupler." *JOSA B* 8.9 (1991): 1914-1931.

[33] Grinblat, Gustavo, et al. "Efficient ultrafast all-optical modulation in a nonlinear crystalline gallium phosphide nanodisk at the anapole excitation." *Science Advances* 6.34 (2020): eabb3123.

[34] Firby, C. J., and A. Y. Elezzabi. "Nonlinear optical modulation in a plasmonic Bi: YIG Mach-Zehnder interferometer." *Integrated Optics: Devices, Materials, and Technologies XXI*. Vol. 10106. International Society for Optics and Photonics, 2017.

[35] Jia, Yue, et al. "Nonlinear optical response, all optical switching, and all optical information conversion in NbSe 2 nanosheets based on spatial self-phase modulation." *Nanoscale* 11.10 (2019): 4515-4522.

[36] Xie, Ze Tao, et al. "Tunable Electro-and All-Optical Switch Based on Epsilon-Near-Zero Metasurface." *IEEE Photonics Journal* 12.4 (2020): 1-10.

[37] Gorbach, A. V. "Nonlinear graphene plasmonics: amplitude equation for surface plasmons." *Physical Review A* 87.1 (2013): 013830.

[38] Hilgevoord, Jan. "The uncertainty principle for energy and time." *American Journal of Physics* 64.12 (1996): 1451-1456.

[39] Goyal, Amit, Thokala Soloman Raju, and C. Nagaraja Kumar. "Lorentzian-type soliton solutions of ac-driven complex Ginzburg–Landau equation." *Applied Mathematics and Computation* 218.24 (2012): 11931-11937.

[40] Golovinski, P. A., A. V. Yakovets, and V. A. Astapenko. "Linear build-up of Fano resonance spectral profiles." *Applied Physics B* 124.6 (2018): 111.

[41] Golovinski, Pavel Abramovich, Andrey Vasil evich Yakovets, and Egor Sergeevich Khramov. "Application of the coupled classical oscillators model to the Fano resonance build-up in a plasmonic nanosystem." *arXiv preprint arXiv:1711.02498* (2017).

[42] Liu, Zhiguang, et al. "Fano resonance Rabi splitting of surface plasmons." *Scientific reports* 7.1 (2017): 1-9.

[43] Ávalos-Ovando, Oscar, et al. "Temporal plasmonics: Fano and Rabi regimes in the time domain in metal nanostructures." *Nanophotonics* 9.11 (2020): 3587-3595.

[44] Boyd, Robert W. *Nonlinear optics*. Academic press, 2020.